\documentclass[a4paper, 12pt]{article}
\usepackage{biblatex}
\usepackage{amsmath,amssymb,amsfonts,amsthm,amscd,makeidx,stackrel,stmaryrd,float,epsfig}
\usepackage[left=2.5 cm,top=2.5 cm,bottom=2.5 cm,right=2.5 cm]{geometry}
\usepackage{amsthm}
\usepackage[english]{babel}
\usepackage[T1]{fontenc}
\usepackage[utf8]{inputenc}
\usepackage{csquotes}
\usepackage{array}
\usepackage{bm}
\usepackage{eucal}
\numberwithin{equation}{section}
\usepackage{graphicx}
\usepackage{subcaption}
\usepackage{here}
\usepackage{float}
\usepackage{url}
\usepackage{colortbl}
\usepackage[export]{adjustbox}
 \usepackage{siunitx}
\usepackage{booktabs,amsmath}
\usepackage{wrapfig}
\usepackage{lipsum}
\usepackage{textcomp}
\usepackage{gensymb}
\usepackage{siunitx}
\usepackage{tikz}
\usetikzlibrary{shapes,arrows}
\usepackage{verbatim}
\usepackage[colorlinks = true,
           linkcolor = blue,
           urlcolor  = blue,
            citecolor =red]{hyperref}

\begin{document}

\begin{center}
\Huge  \textbf{ADCS Preliminary Design For  GNB } 
\end{center}

\begin{center}
\large Alessio Bocci  \footnote{$^\dag$Corresponding Author e-mail: alessiobocci@live.it} $^\dag$, Giovanni Mingari Scarpello \footnote{Second Author e-mail: giovannimingari@yahoo.it}
\end{center}

\begin{abstract}
This work deals with  an ADCS model for a satellite orbiting around Earth. The object is to achieve a preliminary design and perform some analysis on it. To do so, a GNB  was selected and main properties are exploited. Previous works of  \cite{carroll2004arc}, \cite{elliott2014thermal}, \cite{greene2009attitude}, \cite{johnston2011high}  and \cite{philip2008attitude}  were analyzed and a synthesis was obtained; then a suitable control system was designed to satisfy technical requirements. Coding was performed using Matlab and Simulink.

\end{abstract}

\section*{List of Acronyms}

\noindent

\textbf{ADCS}: Attitude Determination and Control System 

\textbf{BRITE}: BRIght Target Explorer

\textbf{ECI}: Earth Centered Inertial (Reference Frame)

\textbf{EPS}: Electrical Power System 

\textbf{FOV}: Field Of View

\textbf{GNB}: Generic Nanosatellite Bus

\textbf{OBDH}: On Board Data Handling 

\textbf{QUEST}: QUaternion ESTimator

\textbf{RMS}: Root Mean Square 

\textbf{RW}: Reaction Wheel

\textbf{SRP}: Solar Radiation Pressure

\textbf{SFL}: Space Flight Laboratory

\textbf{TCS}: Thermal Control System

\textbf{TLE}: Two Line Elements

\textbf{TMTC}: Telemetry and Telecommand (System)

\textbf{TRIAD}: TRIaxial Attitude Determination

\textbf{UTIAS}: University of Toronto Institute for Aerospace Studies

\textbf{mNm}: milli-Newton-meter

\textbf{mNms}: milli-Newton-meter-second

\section{Introduction}

Satellites are roughly consisting of two parts,  the payload and the satellite bus. The payload is responsible for completing the primary objectives of the mission, while the satellite bus supports the payload in its operation (mechanical structure,  EPS, TCS,  OBDH, ADCS, TMTC). 

In order to reduce the development cost of the  bus, the Space Flight Laboratory  \cite{SFL} developed the so called Generic Nanosatellite Bus (GNB) designed to work with a wide array of payloads applying minimal changes to the bus configuration; it is  a highly capable and versatile platform designed to handle a wide variety of missions.

\subsection{BRITE Constellation Mission} 
As reported by \cite{eoportal},  BRITE (BRIght-star Target Explorer)  is a low-cost Austrian/Canadian constellation of nanosatellites born as a collaborative science demonstration mission; the original concept of a  single satellite photometry mission  has grown to a six satellite constellation with science teams, engineering teams and funding sources in Canada, Austria and Poland. 

As detailed in \cite{weiss2014brite}, the aim of the mission is to perform long term stellar observations in order to detect the variations in apparent luminosity of certain stars to understand their history, core composition and internal structure. 

 BRITE mission measurements will do research about the generation of heavy elements, the creation of planets, and the ecology of the Universe.
  
Primary targets are hot luminous stars, massive stars, and stellar supernovae and, as secondary objective, cool luminous stars, such as Red Giants, which are rich in carbon and neutron production. 

\section{Satellite Layout}
\begin{figure}[H]
\centering
\includegraphics[scale=0.5]{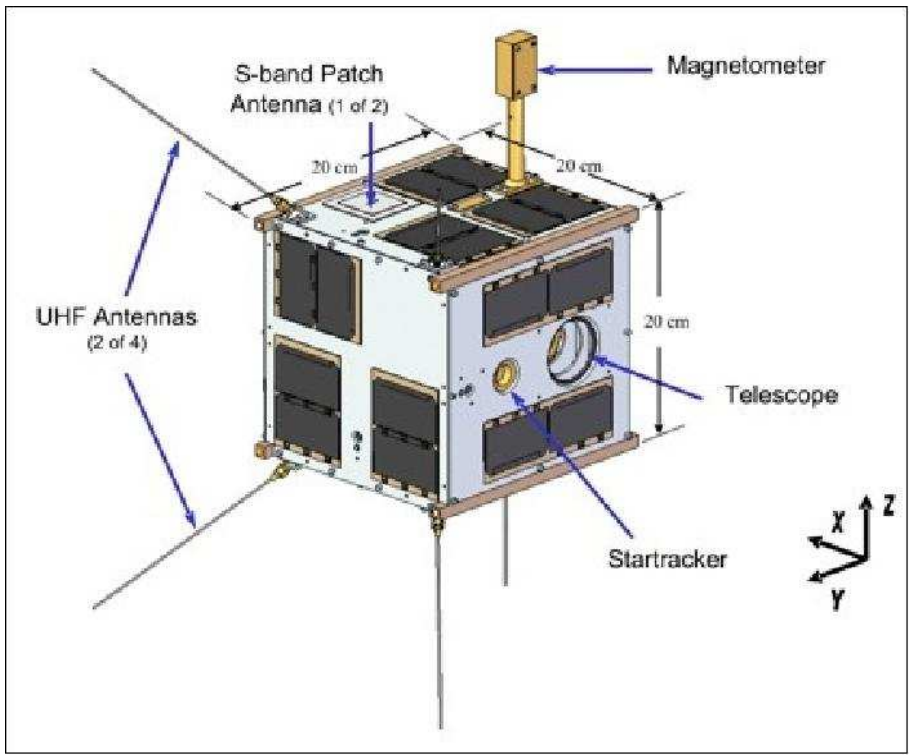}
\caption{BRITE Configuration (courtesy of \cite{eoportal})}
\label{BRITE1}
\end{figure}

\begin{figure}[H]
\centering
\includegraphics[scale=0.5]{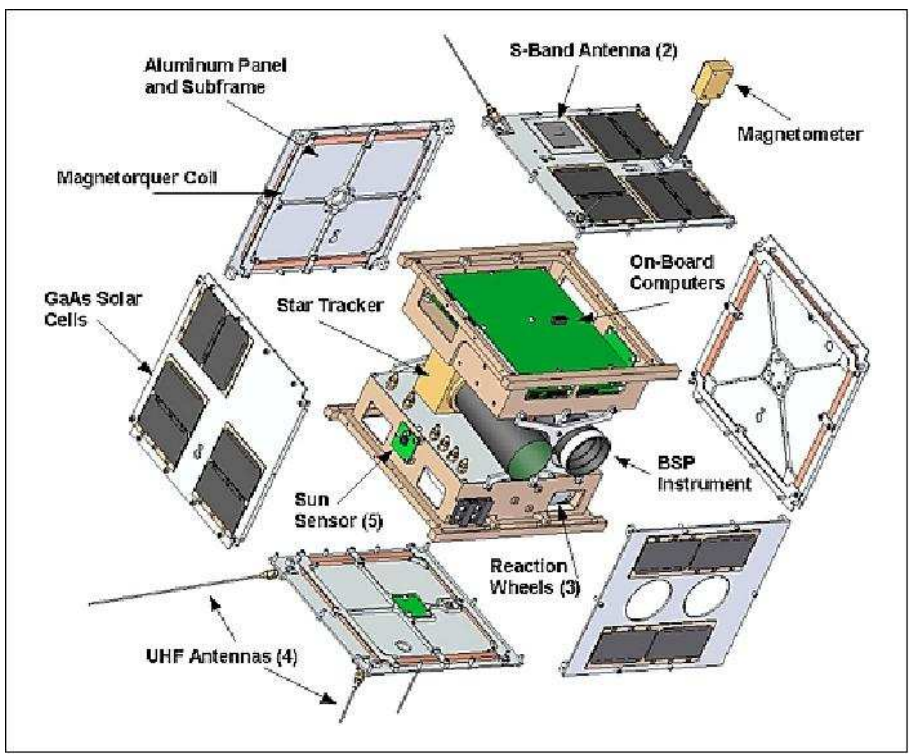}
\caption{BRITE Configuration - Exploded View (courtesy of \cite{eoportal})}
\label{BRITE2}
\end{figure}

The typical configuration of the BRITE satellite is presented in Figure~\ref{BRITE1} and Figure~\ref{BRITE2}, essentially it is a 20 cm x 20 cm x 20 cm cube with a mass of 7 Kg.

The 2 mm thick panels with cross braces provide additional rigidity to the structure supporting a mounting surface for the solar cells, most  sun sensors, the magnetometer boom, the magnetorquers and the various antennas. All other components, such as the reaction wheels, various computers on board, battery assemblies, radios and some of the sun sensors are mounted directly to the traies.

In \cite{carufel2009assembly},\cite{elliott2014thermal},\cite{greene2009attitude} and \cite{johnston2011high}   all components, instrumentation (in particular sensors and actuators for ADCS) and geometrical/mass properties are analyzed and assembly and integration procedures are presented. In \cite{johnston2011high} its  Inertia Tensor is reported as:

\begin{equation}
J=\left[ \begin{array}{ccc}
0.0465&- 0.0007&0.0004 \\
   -0.0007& 0.0486& -0.0021\\
    0.0004& -0.0021& 0.0482
\end{array}\right] \quad \text{Kg}\cdot \text{m}^2
\end{equation}

\section{Environment of the Mission and Mathematical Models }

\subsection{Orbital Motion}

From \cite{nasa}, \cite{space} and \cite{cele}  and  is possible evaluate the spacecraft TLEs from his catalog number. In our case the NORAD CATALOG NUMBER is 40020 and the following orbit properties are reported in Table~\ref{OrbProp}.

\begin{table}[H]
\centering
\begin{tabular}{|c|c|c|c|c|}
\hline
\textbf{Perigee Altitude} & \textbf{Apogee Altitude} & \textbf{Period} & \textbf{Eccentricity} & \textbf{Inclination}\\
\hline
$611$ Km & $732$ Km & $98.18$ min & 0.0085944 &$\ang{97.73}$\\
\hline 
\end{tabular}
\caption{Orbit Properties}
\label{OrbProp}
\end{table}
Therefore from TLEs is possible to evaluate the initial conditions in terms of position and velocity for orbit propagation.
Denoting with $\bm{r}=[x,y,z]^T$ the position vector of the spacecraft in the ECI reference frame, the restricted two body problem equation is:
\begin{equation}
\ddot{\bm{r}}+\frac{\mu}{||\bm{r}||^3}\cdot \bm{r}=\bm{a}_p
\label{2bp}
\end{equation}
where  $\mu$ is the Earth gravitational constant. 

The  $\bm{a}_p$ represents the perturbative accelerations acting on the spacecraft; we considered:
\begin{enumerate}
\item Perturbation due to Earth oblateness: $J_2$ term (see \cite{curtis2013orbital} page 491 equation 10.30)

\item Perturbation due to Drag: Earth atmosphere effect (see \cite{markley2014fundamentals} page 108 equation 3.163)

\item Perturbation due to Solar Radiation Pressure: SRP (see \cite{markley2014fundamentals} page 109 equation 3.167)

\item Perturbation due to Third Body: Moon (see \cite{curtis2013orbital} page 530 equation 10.117) 
\end{enumerate}

It is important to underline that as far as SRP and Drag are concerned, the force is directly correlated to the area seen by the relative velocity vector and therefore is strictly connected to the attitude that the satellite is assuming in any instant of time.
We assumed  the value of the Drag Coefficient is $C_d=2.6$. 
The eclipse condition is taken into consideration through the cylindrical approximation:

\begin{equation}
\bm{r}\times \bm{e}_{e-s}<- \sqrt{||\bm{r}||^2-R_e^2}
\label{eclipse}
\end{equation}

where $\bm{e}_{e-s}$ is the unit vector from the Earth to the Sun and $R_e$ is the Earth radius. 

The $J_2$ effect is negligible for short time mission operations, but as long as time passes, its effect becomes evident; in fact the orbit drifting implies the changing in keplerian elements and in particular on the inclination of the orbit. Therefore the Earth magnetic field sensed by the spacecraft will be strongly affected by it.

\subsection{Attitude Motion}
The mathematical model adopted is based on Euler's equations of motion, denoting with $\bm{\omega}$ the body's angular velocity vector and with $J$ its inertia tensor, we have:

\begin{equation}
J \dot{\bm{\omega}}=(J\bm{\omega})\wedge \bm{\omega} + \bm{M}_d+\bm{L}
\label{euler}
\end{equation}

In \eqref{euler} $\bm{L}$ represents the control torque (we will see later how to define it according to the mission purposes) while $\bm{M}_d$ is the disturbance torque. We considered the following disturbances:

\begin{enumerate}
\item Gravity Gradient torque (see \cite{markley2014fundamentals} page 104 equation 3.155)
\item Drag Torque (see \cite{markley2014fundamentals} page 108 equation 3.164)
\item Solar Radiation Pressure Torque (see \cite{markley2014fundamentals} page 109 equation 3.168)
\item Magnetic Torque  (see \cite{markley2014fundamentals} page 405 equation 11.5 for the dipole approximation for the magnetic field and page 106 equation 3.159 for the magnetic torque)
\end{enumerate}

Once \eqref{euler} is integrated, the kinematics in terms of quaternion $\bm{q}$ is simply:
\begin{equation}
\dot{\bm{q}}=\frac{1}{2} U(\bm{\omega}) \bm{q}
\end{equation}
where:
\begin{equation}
U(\bm{\omega})=\left[\begin{array}{cccc}
0 & \omega_3 & -\omega_2 & \omega_1\\
-\omega_3 & 0 & \omega_1 &\omega_2 \\
\omega_2&-\omega_1 &0 &\omega_3 \\
-\omega_1 & -\omega_2 & -\omega_3 &0
\end{array} \right]
\end{equation}

\section{Sensors and Actuators}
For attitude determination the typical GNB (see \cite{greene2009attitude}) configuration is:
\begin{enumerate}
\item \textbf{6 Sun Sensors} (one on each face): 
unit sphere coverage of  $99.8 \%$. Main properties are reported in Table~\ref{sussens}.
\begin{table}[H]
\centering
\begin{tabular}{|c|c|c|c|c|}
\hline
\textbf{Mass }& \textbf{Dimensions }& \textbf{Accuracy} & \textbf{FOV} & \textbf{Resolution} \\
\hline
$<6$ g & $30$ mm x $30$ mm & $<\ang{1.25}$ RMS & $\ang{90.62}$ (min) $-$ $\ang{107.92}$ (max)  & $\ang{0.422}$/pixel\\ 
\hline
\end{tabular}
\caption{Sun Sensors Main Properties}
\label{sussens}
\end{table}
\item \textbf{1 Magnetometer}: it is structurally located over the body, in order to ensure minimal measurement corruption by the residual magnetic dipole of the satellite. This sensor  can provide attitude knowledge in the range of $\pm \ang{1}$.

\item \textbf{1 Star Tracker}:   Miniature Star Tracker (MST) developed by AeroAstro  which meets the accuracy requirements of the BRITE mission while fitting into the volume of only one quarter of the GNB payload bay. Main properties are reported in Table~\ref{starssens}.

\begin{table}[H]
\centering
\begin{tabular}{|c|c|c|c|c|}
\hline
\textbf{Mass }& \textbf{Dimensions }& \textbf{Accuracy} & \textbf{Power} & \textbf{Stars Tracked} \\
\hline
$375$ g & $6$ cm x $7.62$ cm x $7.62$ cm& $\pm 70$ arc seconds & $<2$ W & $>9$\\ 
\hline
\end{tabular}
\caption{Star Tracker Main Properties}
\label{starssens}
\end{table}

\end{enumerate}

For attitude control the typical GNB configuration is:
\begin{enumerate}
\item \textbf{3 Orthogonal Reaction Wheels}: the Sinclair/SFL \cite{sinclair}  reaction wheels are the primary control actuator for all GNBs and are responsible of slewing the spacecraft bus to a desired attitude. Main properties are reported in Table~\ref{rw}.

\begin{table}[H]
\centering
\begin{tabular}{|c|c|c|c|c|}
\hline
\textbf{Mass }& \textbf{Dimensions }& \textbf{Moment of } & \textbf{Momentum} & \textbf{Maximum} \\
&&\textbf{Inertia} & \textbf{Capacity}& \textbf{Torque}\\
\hline
$185$ g & $5$ cm x $5$ cm x $4$ cm& $5.12 \cdot 10^{-5}$ $\text{Kg}\cdot\text{m}^2$ & $30$ mNms & $2$ mNm\\ 
\hline
\end{tabular}
\caption{Star Tracker Main Properties}
\label{rw}
\end{table}

\item \textbf{3 Orthogonal Magnetorquers}:  are the second type of control actuator used on the GNB; they are used for the detumbling phase and for reaction wheels desaturation. Main properties are reported in Table~\ref{magnetorquers}.

\begin{table}[H]
\centering
\begin{tabular}{|c|c|c|c|}
\hline
\textbf{Mass }& \textbf{Dimensions }& \textbf{Power Consumption } & \textbf{Dipole Moment} \\
\hline
$30$ g & $8$ cm x $8$ cm x $10$ cm& $300$ mW & $0.12$ A$\text{m}^2$ \\ 
\hline
\end{tabular}
\caption{Magnetorquers Main Properties}
\label{magnetorquers}
\end{table}
\end{enumerate}

\section{Mission Requirements}
The ADCS requirements can be formalized as:

\begin{itemize}
\item The ADCS shall allow imaging of a target field for 15 minutes continuously for each orbit, with a goal of imaging multiple targets each orbit
\item The ADCS should accommodate observations of multiple targets in a single orbit
\item The ADCS shall provide a pointing accuracy less then $\ang{1}$ (1 arcmin RMS)  on each target 
\item The Attitude Determination shall require less then 10 arcsec
\item The ADCS shall be capable of operating for the full duration of the mission (2 years)
\end{itemize}

\section{Main Quantities Estimation and Filtering}

\subsection{Attitude Determination}
As already mentioned, the spacecraft is equipped with various sensors for attitude determination, whose features and operating mechanisms are discussed in detail in \cite{wertz2012spacecraft}. 

 Our implementation logic  is to use the pair sun sensors - magnetometer for attitude determination in the initial phase of detumbling in order  to give to the star tracker enough time to start correctly and in case of failure of the latter.
 
This combination of sensors is also used for  avoiding the  Sun in order not to damage the star tracker.

 In all other cases, the attitude determination is performed by the star tracker. Therefore this is a redundant configuration.
 
 When using Sun sensors and magnetometer, the estimation is less precise than star tracker one; however, if the spacecraft is not in eclipse condition, they provide two measurements capable  of  determining the attitude accordingly, in order to solve the  sun sensors - magnetometer attitude determination problem, the  static  algorithm  TRIAD has been chosen.   This method is easy and no computational problems are present.

 On the other hand the star tracker provides much more measurements at the same time; in fact we need at least eight stars to have $90 \%$ of probability to find four stars in the FOV. 
 
 To be more realistic, we decided to build a star catalog in which 
we considered eight stars; using  their right ascension and declination, director cosines of the star position vector can be evaluated. 

In order to model the FOV limitation  a sort  "eclipse condition"  \eqref{eclipse} is adopted. The algorithm logic is that of determine the stars in the FOV, then remove the other for the next calculations to reduce the computational cost.

The  chosen static determination algorithm in this case is the QUEST method instead of Davenport's q method since it is more precise but more computationally expansive. 

\subsection{Angular Velocity Estimation}

Since no \emph{rate sensors} are present the angular velocity $\bm{\omega}$ has to be estimated ex post (this problem is also analyzed by \cite{cortiella20163cat}).
The quaternion kinematics can be described by:
\begin{equation}
\dot{\bm{q}}= \frac{1}{2}\Xi(\bm{q})\bm{\omega} \quad 
\quad \text{with}\quad
\Xi(\bm{q})=\left[ \begin{array}{ccc}
q_4 &-q_3 &q_2\\
q_3 &q_4&-q_1\\
-q_2&q_1&q_4\\
-q_1&-q_2&-q_3
\end{array}\right]
\label{quat1}
\end{equation}

Equation \eqref{quat1} can be then inverted as:
\begin{equation}
\bm{\omega}=2 \Xi^T(\bm{q})\dot{\bm{q}}
\end{equation}
Then the derivative  $\dot{\bm{q}}$ of the quaternion can be evaluated using finite difference method. 

In a same fashion the Attitude Matrix $A_{B/N}$ that represent the map from the inertial frame ($N$) to the body ($B$) frame (direction cosines matrix) can be computed as:
\begin{equation}
\dot{A}_{B/N}=-[\bm{\omega}\wedge]A_{B/N} \rightarrow [\bm{\omega}\wedge]=-\dot{A}_{B/N}A_{B/N} ^T
\end{equation} 
where:
\begin{equation}
[\bm{\omega}\wedge]=\left[ \begin{array}{ccc}
0 &-\omega_3 & \omega_2 \\ \omega_3 & 0 &-\omega_1\\ -\omega_2 & \omega_1&0
\end{array}\right]
\end{equation}

and finally we have:
\begin{equation}
\bm{\omega}=(-\dot{A}_{B/N}A_{B/N}^T )^V, \quad \text{with} \quad ( P)^V=[P_{32},P_{13},P_{21}]^T
\label{Vop}
\end{equation}
In \eqref{Vop} P is a generic matrix. 
Then the derivative $\dot{A}_{B/N}$ of the Attitude Matrix   can be evaluated using finite difference method. 

\subsection{Filtering}
The TRIAD algorithm and the QUEST method provide a good approximation of quaternion and attitude matrix despite the random error related to the inaccuracy of the sensors. 

On the other hand, the numerical derivative generates   noisy signals that have to be filtered.  In order to eliminate high frequency noise without loosing too much information from the signal, two different low pass filter are implemented whose transfer functions are:

\begin{equation}
F_1(s)=\frac{\tilde{\omega}_1}{s+\tilde{\omega}_1}, \quad F_2(s)=\frac{\tilde{\omega}_2^2}{1+2 \xi \tilde{\omega}_2 s+\tilde{\omega}_2^2}
\end{equation}  
where  $\tilde{\omega}_1$  and $\tilde{\omega}_2$ are the cut off frequencies of the first order and second order low pass filter respectively and $\xi$ is the damping ratio of the second order low pass filter.

All these parameters are optimally tuned in order to obtain the best performances depending on each mission operation  phase.

\section{Detumbling Maneuver}
The detumble mode shall reduce the satellite rotation rates after release from the launcher,  and after entering safe-hold mode,  to the point where satellite angular momentum is low enough to be absorbed by reaction wheels.  For BRITE,  this is designed to recover from tumble rates up to 60 deg/s.

The control law is chosen as a combination of a bang-bang control  and a pure B-dot control law \footnote{This is a simplified magnetic control law, more advanced ones are present in literature such as  \cite{rodriquez2015spacecraft}}:

\begin{equation}
\bm{L}_{det}= 
\begin{cases}
 -\bm{b} \wedge \bm{m}    & \quad \text{if} \quad ||\bm{\omega}|| \geq 1 \deg/s \\
 -k_{det} (\bm{b} \wedge \bm{\omega})\wedge \bm{b}& \quad \text{if} \quad ||\bm{\omega}||<1 \deg/s \
\end{cases}
\label{detumbling}
\end{equation} 

Where: 

\begin{equation}
\bm{b}=\dfrac{\bm{B}}{||\bm{B}||} , \quad  \bm{m}=-m_{max} \cdot \text{sign}(\dot{\bm{b}}) \quad \text{and} \quad  k_{det}=\dfrac{4 \pi}{T_{orb}}[1+\sin(\xi_m)]J_{min}
\label{par1}
\end{equation}

Where $\bm{B}$ is the Earth magnetic field vector, $\bm{m}$ is the dipole moment of the magnetorquer, $m_{max}$ is the maximum magnetic dipole of the magnetorquer, $T_{orb}$ is the orbital period, $\xi_m$ is the inclination of the spacecraft orbit relative to the geomagnetic equatorial plane and $J_{min}$ is the minimum principal moment of inertia.
The formula for $k_{det}$ in \eqref{par1} can be found in \cite{markley2014fundamentals} and a detailed proof is in \cite{avanzini2012magnetic}. Typical values of $m_{max}$ for GNB satellites are presented in \cite{carroll2004arc} , in our case $m_{max}=0.12$ $A m^2$. 

The satellite is equipped with one star tracker,  six sun sensors and one magnetometer, therefore the implementation logic is such that the attitude sensing is performed using sun sensors and the magnetometer for at least 60 minutes (which more or less correspond with the time the satellite is not in eclipse) end possibly,  if needed,  the star tracker performs the remaining of the attitude sensing to conclude the detumbling.  This strategy is done in order to give enough time to the star tracker to switch on correctly (maximum time 60 minutes).
   
   \begin{figure}[H]
   \centering
   \includegraphics[scale=0.3]{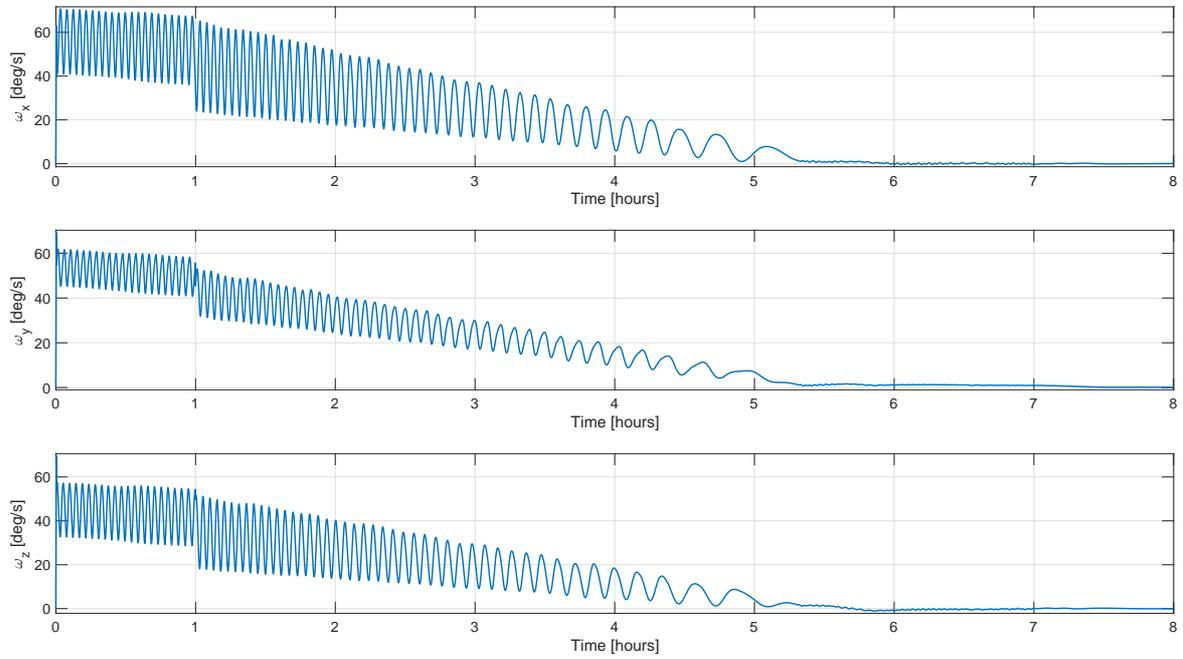}
   \caption{Detumbling Simulation Results - Angular Velocity Components Behaviour}
   \label{OmegaDetumbling}
   \end{figure}

Figure~\ref{OmegaDetumbling} shows the behaviour of the angular velocity components,  the initial condition $\bm{\omega}(0)$ is selected as the limit one; we can see that the asymptotic zero condition is achieved more or less in 6 hours.
   
   \begin{figure}[H]
   \centering
   \includegraphics[scale=0.3]{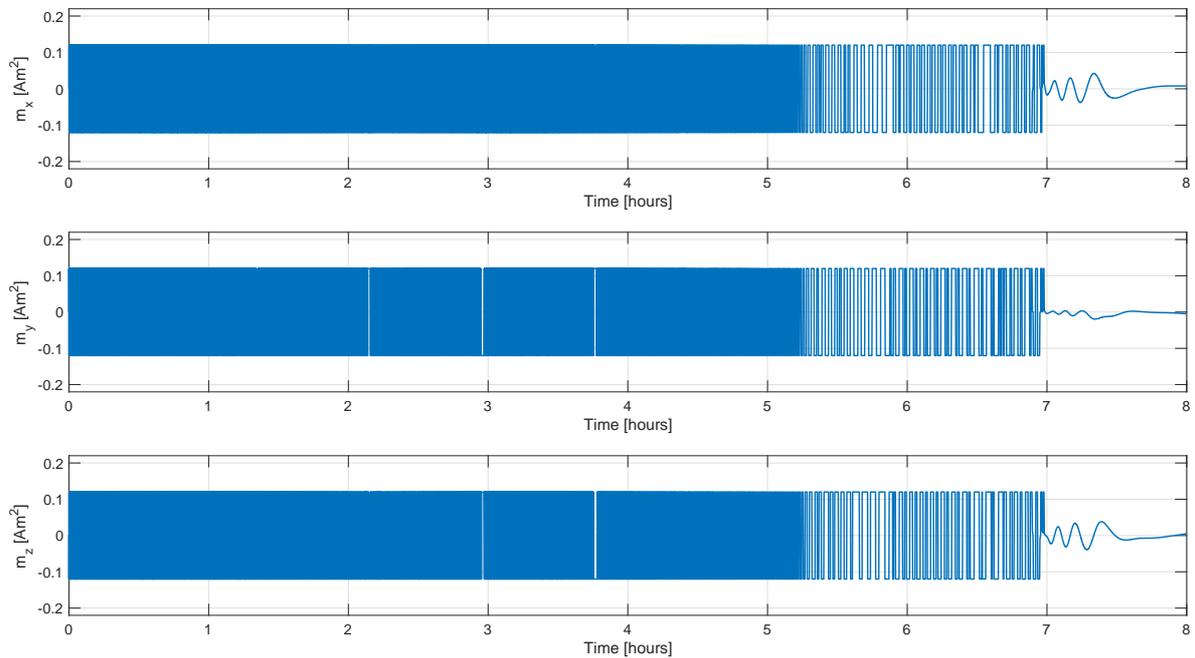}
   \caption{Detumbling Simulation Results - Dipole moment components}
   \label{DipoleDetumbling}
   \end{figure}
   
      \begin{figure}[H]
   \centering
   \includegraphics[scale=0.3]{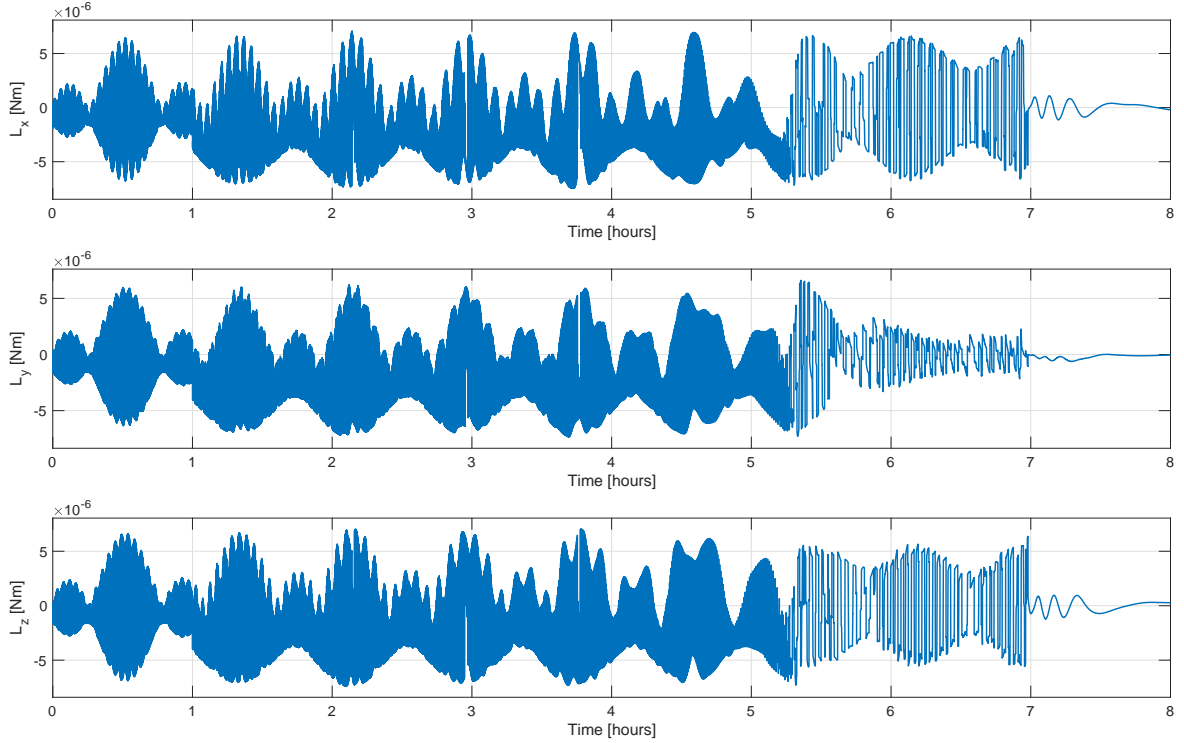}
   \caption{Detumbling Simulation Results - Control Torque components}
   \label{TorqueDetumbling}
   \end{figure}
   
Figure~\ref{DipoleDetumbling} presents the behaviours of magnetic dipole $\bm{m}$ components; here we can appreciate the difference and the switching of  two methods.  Figure~\ref{TorqueDetumbling} presents the required torque, called $\bm{L}$ for simplicity.
Such a simulation,  with so onerous initial conditions,  shows also the robustness of the control design; in fact it is capable of detumbling successfully the spacecraft despite the long amount of time required. Therefore less onerous initial conditions will require less effort.

\section{Slew and Tracking Maneuvers}   
After the detumbling maneuver,  the spacecraft shall point to the target. For our purposes, the goal is to point a fixed star.  Two main consequences can be identified: the \emph{reference attitude} (desired attitude) is not time varying with respect the ECI reference frame and the \emph{desired angular velocity} during the tracking is zero. 
Therefore we are allowed to formulate the tracking and slew problem on the same time as:
\begin{equation}
\bm{L}_{s-t}=-k_\omega \bm{\omega}+\bm{\omega} \wedge (J \bm{\omega})-k_A(A_e^T-A_e)^V
\label{trackinglaw}
\end{equation}
Where:
\begin{equation*}
(A_e^T-A_e)^V=[\hat{A}_{32}, \hat{A}_{13}, \hat{A}_{21}]^T, \quad \text{with} \quad  \hat{A}=A_e^T-A_e
\end{equation*}

Equation \eqref{trackinglaw} can be proved as follows:
\begin{proof}
Let $\bm{\omega}_e=\bm{\omega}-A_e\bm{\omega}_d$ the error in angular velocity with respect to the desired set point $\bm{\omega}_d$.
The condition of equilibrium to be studied is $\bm{\omega}_e=\bm{0}$, 
then a  suitable Lyapunov function is:
\begin{equation}
V(\bm{\omega}_e)=\frac{1}{2}\bm{\omega}_e^T J\bm{\omega}_e+k_A\text{tr}(I-A_e)
\label{pip}
\end{equation}
which is zero for $\bm{\omega}_e=\bm{0}$ and $A_e=I$. The chosen Lyapunov function is never less than zero because the first term is a quadratic form and the trace surely greater than zero.
Performing the time derivative of \eqref{pip} we have:
\begin{align*}
\dot{V}(\bm{\omega}_e)&=\bm{\omega}_e^T J\dot{\bm{\omega}}_e+k_A \text{tr}([\bm{\omega}\wedge]A_e)\\
&=\bm{\omega}_e^T \left[(J\bm{\omega})\wedge \bm{\omega}+\bm{L}-J A_e \dot{\bm{\omega}}_d+J[\bm{\omega}_e\wedge]A_e\bm{\omega}_d+k_A(A_e^T-A_e) \right]
\end{align*}
Where the external disturbances are not considered since in general it is not possible to compensate them. 
Finally we derive the general control law:
\begin{equation}
\bm{L}=-k_\omega \bm{\omega}_e+\bm{\omega}\wedge(J\bm{\omega})+J(A_e \dot{\bm{\omega}}_d-[\bm{\omega}_e\wedge]A_e\bm{\omega}_d)-k_A(A_e^T-A_e)^V
\label{pluto}
\end{equation}
which has been chosen in such a way that $\dot{V}(\bm{\omega}_e)<0$.

By setting $\bm{\omega}_d=\bm{0}$ in \eqref{pluto} we get \eqref{trackinglaw} and the system is asymptotically stable.
\end{proof}

From \eqref{trackinglaw} we can see that first of all we perform a feedback linearization by means of the term $\bm{\omega} \wedge (J \bm{\omega})$ to the Euler's equations and then two proportional actions are applied: one $k_\omega$  on the angular velocity  and the other $k_A$ on the attitude error  ($\{k_\omega,\quad k_a\}>0$).

Equation \eqref{trackinglaw} represents the so called desired control law, then we map it onto actuators solving the following differential equation:

\begin{equation}
\dot{\bm{h}}_{rw}=- \tilde{A}^* [\bm{L}_{s-t}+\bm{\omega} \wedge (\tilde{A}\bm{h}_{rw})]
\label{hrwd}
\end{equation}
where $\tilde{A}$ is the distribution matrix of the momentum exchange devices and $\tilde{A}^*$ its pseudo-inverse (since we have three orthogonal reaction wheels $\tilde{A}=\tilde{A}^*=I\in \mathbb{R}^{3\times 3}$), $\bm{h}_{rw}$ is the angular momentum vector of the reaction wheels.
Equation \eqref{hrwd} can be proven as follows:
\begin{proof}
The total angular momentum of the spacecraft can be expressed as:
\begin{equation}
\bm{h}_{tot}=J \bm{\omega}+\tilde{A}\bm{h}_{rw}
\label{pp}
\end{equation}
performing  time derivative of  \eqref{pp} and remembering that matrix $J$ and $\tilde{A}$ are not time varying, we have:
\begin{equation}
J \dot{\bm{\omega}}+\bm{\omega}\wedge (J \bm{\omega})+\tilde{A} \dot{\bm{h}}_{rw}+\bm{\omega} \wedge (\tilde{A}\bm{h}_{rw})=\bm{L}+\bm{M}_d
\end{equation}
Then \eqref{hrwd} is obtained inverting the relation $\bm{L}=-\tilde{A} \dot{\bm{h}}_{rw}-\bm{\omega} \wedge (\tilde{A}\bm{h}_{rw})$
\end{proof}

Without any loss of generality, we considered as target the star Alpha Circini (a variable star in the faint, southern, circumpolar constellation of Circinus) that has an \emph{apparent magnitude} of 3.18 - 3.21 (that was also imaged and studied by BRITE Austria).  

Alpha Circini's properties with respect to the ECI reference frame are:

\begin{equation}
r_{cir-sun}= 54 \quad \text{ly}, \quad \alpha_{cir}=14h \quad 42m \quad 30.41958s,  \quad \delta_{cir}=- \ang{64} \quad 58'  \quad 30.4934''
\label{info}
\end{equation}

where $ r_{cir-sun}$ is the distance from the Sun,  $\alpha_{cir}$ is the Right Ascension and $\delta_{cir}$ is the declination of the star.
Therefore from the information in \eqref{info} we can easily compute the position vector with respect to the ECI reference frame.  

If we now consider the unit vector of the x-axis of the ECI reference frame $\bm{i}=[1,0,0]^T$ than the quaternion to rotate $\bm{i}$ into the direction of $\bm{r}_{cir-sun}$ can be evaluated as:

\begin{equation}
\bm{q_d}=[\cos(\psi/2), \bm{u}\sin(\psi/2)]^T, \quad \bm{u}= \frac{\bm{i} \wedge\bm{r}_{cir-sun}}{||\bm{i} \wedge\bm{r}_{cir-sun}||}, \quad \psi=\cos^{-1} \left( \frac{\bm{i} \times\bm{r}_{cir-sun}}{||\bm{r}_{cir-sun}||}\right)
\label{desiredquaternion}
\end{equation}

From \eqref{desiredquaternion} the desired attitude matrix and the attitude error ($A_e=A_{B/N}A_d^T$) can be evaluated by the usual formulas for converting a quaternion into attitude matrix. 

Finally we can evaluate the most important performance parameter for this mission phase,  the so called pointing error that we define as:

\begin{equation}
\theta_e= \cos^{-1} \left(\frac{\text{tr}(A_e)-1}{2} \right)
\label{pointingerror}
\end{equation}  

Equation \eqref{pointingerror} provides a criterion for deciding if we are or not satisfying the pointing requirement.

The initial conditions for the slew maneuver are those at the end of the detumbling one both in terms of angular velocity and attitude quaternion.

In our design is of course necessary to provide a desaturation mechanism for the reaction wheels; this is done by magnetotorquers.   If we define $M_{max}$ the maximum torque that can be produced by a reaction wheel then the torque to be applied to the RW and the torque to be produced by the magnetorquers to be applied to the spacecraft are:

\begin{equation}
\bm{T}_{rw}=-M_{max} \cdot \text{sign}(\bm{h}_{rw}) \quad  \text{and}\quad \bm{L}_{des}=\bm{m}_{des} \wedge \bm{B}, \quad \bm{m}_{des}=M_{max}\cdot\frac{\text{sign}(\bm{h}_{rw})\wedge \bm{b}}{||\bm{B}||}
\label{satur}
\end{equation}
 
 Where $M_{max}$ is the maximum torque that reaction wheels can provide.
Equation \eqref{satur} provides the desaturation mechanism and it is effective if the magnetic field is orthogonal to the spin axis of the wheel; this yields to the possibility of making the desaturation when we reach such a condition preventing the saturation before it actually happens.

 \begin{figure}[H]
 \centering
 \includegraphics[scale=0.3]{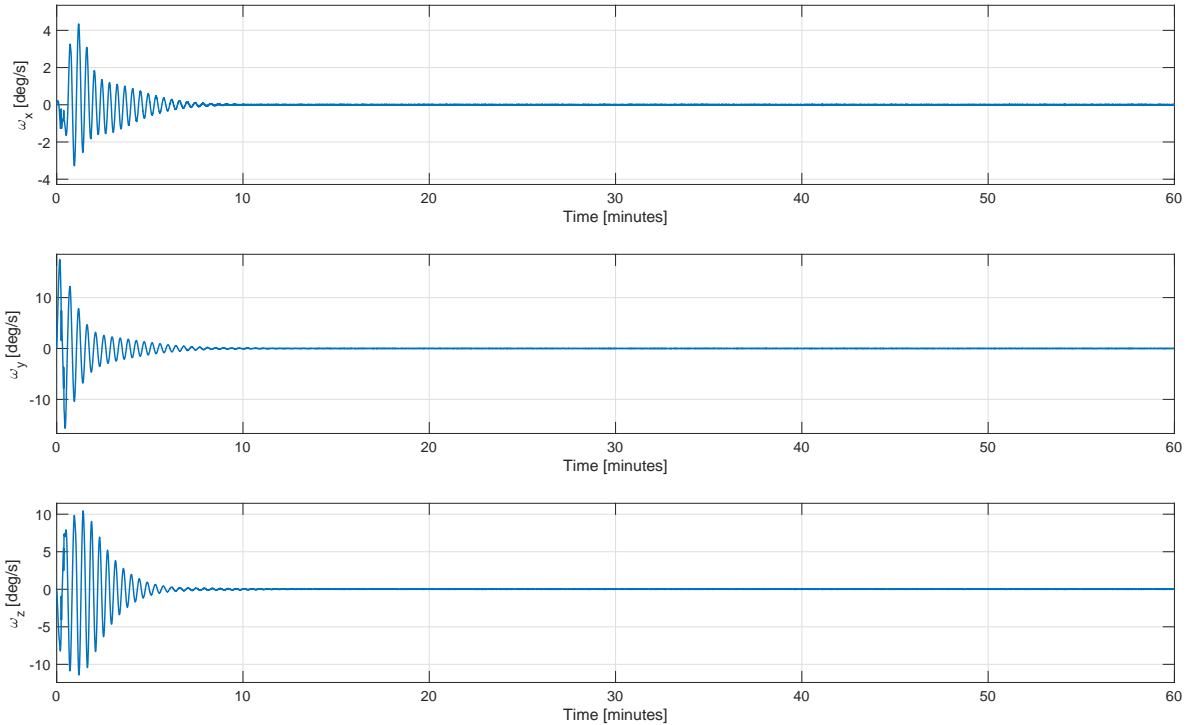}
 \caption{Slew and Tracking Simulation Results - Angular Velocity Components Behaviour}
 \label{OmegaSlew}
 \end{figure}

 \begin{figure}[H]
 \centering
 \includegraphics[scale=0.3]{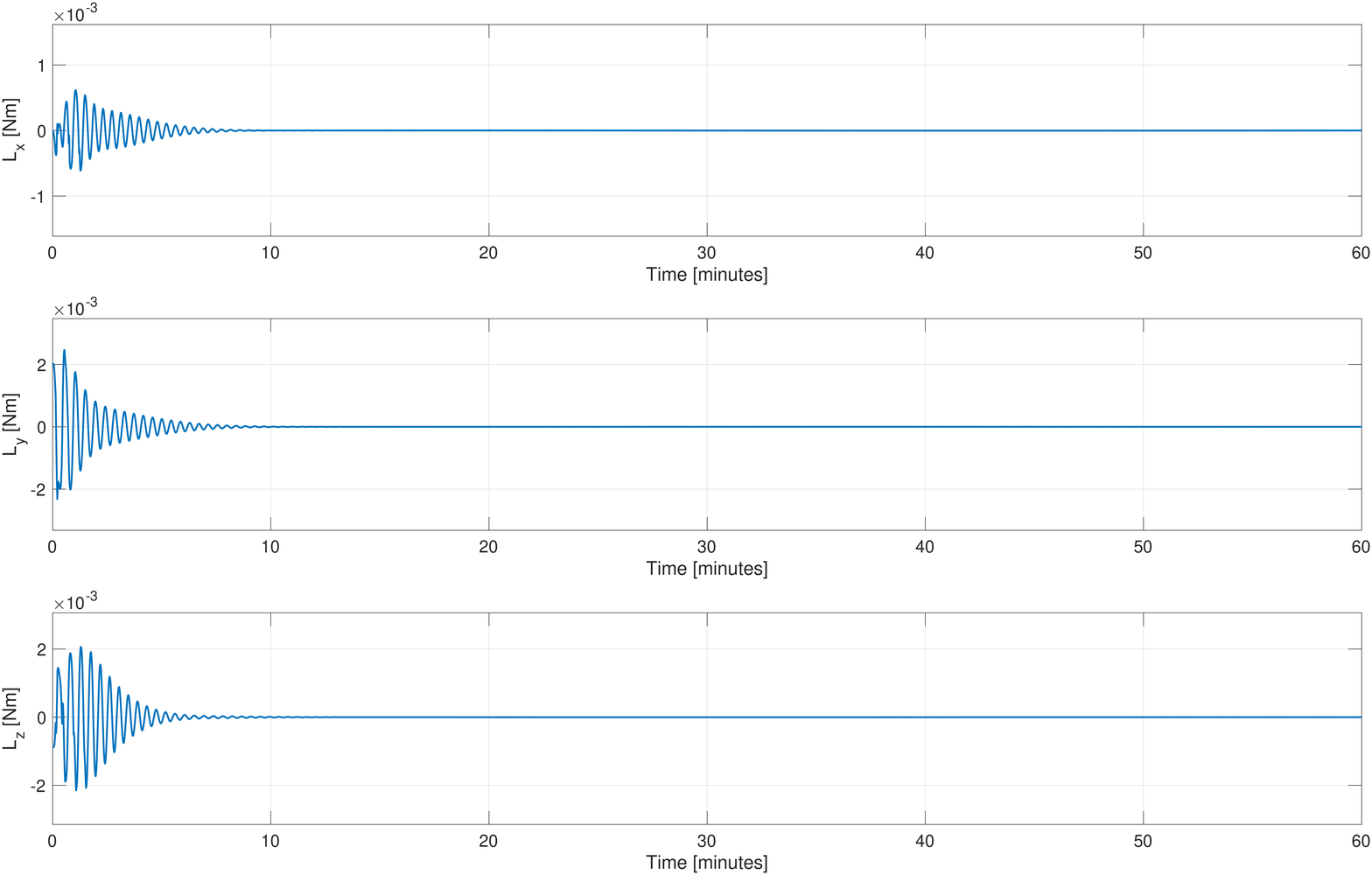}
 \caption{Slew and Tracking Simulation Results - Required Torque}
 \label{TorqueSlew}
 \end{figure}
 
   \begin{figure}[H]
   \centering
   \includegraphics[scale=0.3]{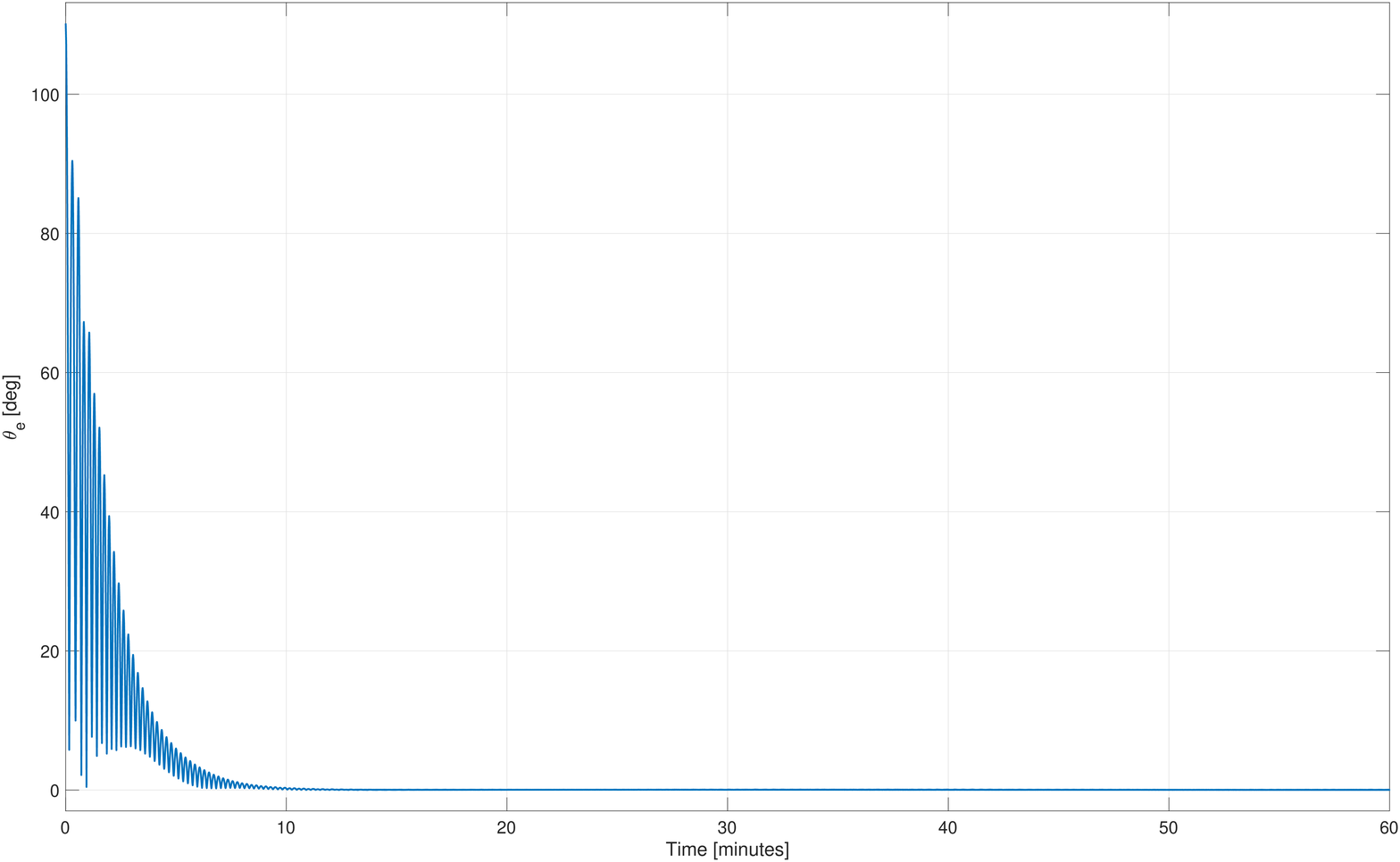}
   \caption{Slew and Tracking Simulation Results - Pointing Accuracy}
   \label{PointingAccuracy}
   \end{figure}
   
   \begin{figure}[H]
   \centering
   \includegraphics[scale=0.3]{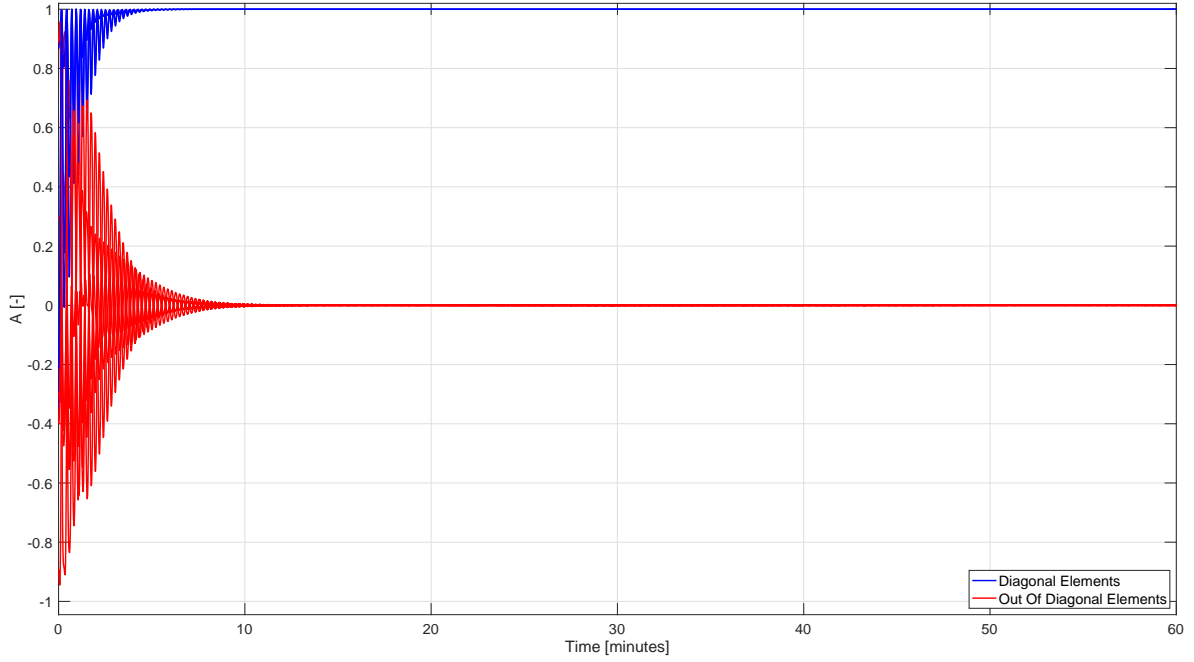}
    \caption{Slew and Tracking Simulation Results - Attitude Error}
    \label{Attitude Error}
   \end{figure}
   
  Figure~\ref{OmegaSlew} and Figure~\ref{TorqueSlew} show respectively the behaviour of the angular velocity and torque components (in   Figure~\ref{TorqueSlew} the torque is labelled as $L$ and not $L_{s-t}$ as in \eqref{trackinglaw} for simplicity); as expected $\bm{\omega}$ start with the zero condition achieved from the detumbling and then start to increase (thanks to the RW actions) to line up the camera with the target direction.
   
   In Figure~\ref{PointingAccuracy}  we can see how the pointing error decreases till the required pointing accuracy $< \ang{1}$ (in our calculation $\approx\ang{0.023}$), this means that the attitude error  tends to the identity matrix as shown in Figure~\ref{Attitude Error}.
 
 These calculations are done considering the saturation and de-saturation mechanism according to \eqref{satur},  however with the designed control the reaction wheels are more or less working  under the saturation limit.

 \section{Re-Pointing}
 
The technical requirements of the ADCS subsystem establish that during one orbital period the satellite must orient the camera towards more than one star and keep this direction for at least 15 minutes (multiple targets each orbit). 

Let us consider as the first target Alpha Circini and as second target Tania Australis.

 Tania Australis is a binary star in the constellation of Ursa Major. An apparent visual magnitude of +3.06 places it among the brighter members of the constellation. This star has the following properties with respect to the ECI reference frame are:
 
\begin{equation*}
r_{tania-sun}= 230 \quad \text{ly}, \quad \alpha_{tania}=10h \quad 22m \quad 19.73976s,  \quad \delta_{tania}=\ang{41} \quad 29'  \quad  58.2691''
\end{equation*}
 
\begin{figure}[H]
 \centering
 \includegraphics[scale=0.3]{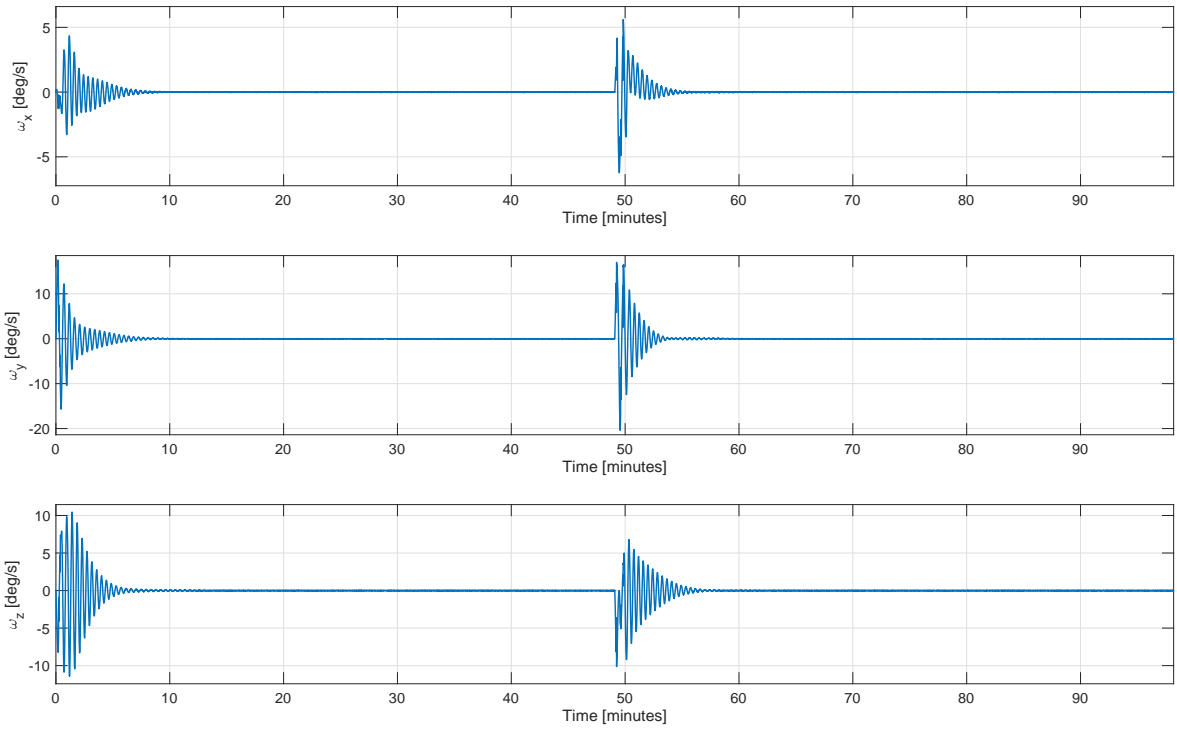}
 \caption{Slew and Tracking Simulation Results - Angular Velocity Components Behaviour (Re-pointing)}
 \label{OmegaSlewRepointing}
 \end{figure}

\begin{figure}[H]
 \centering
 \includegraphics[scale=0.3]{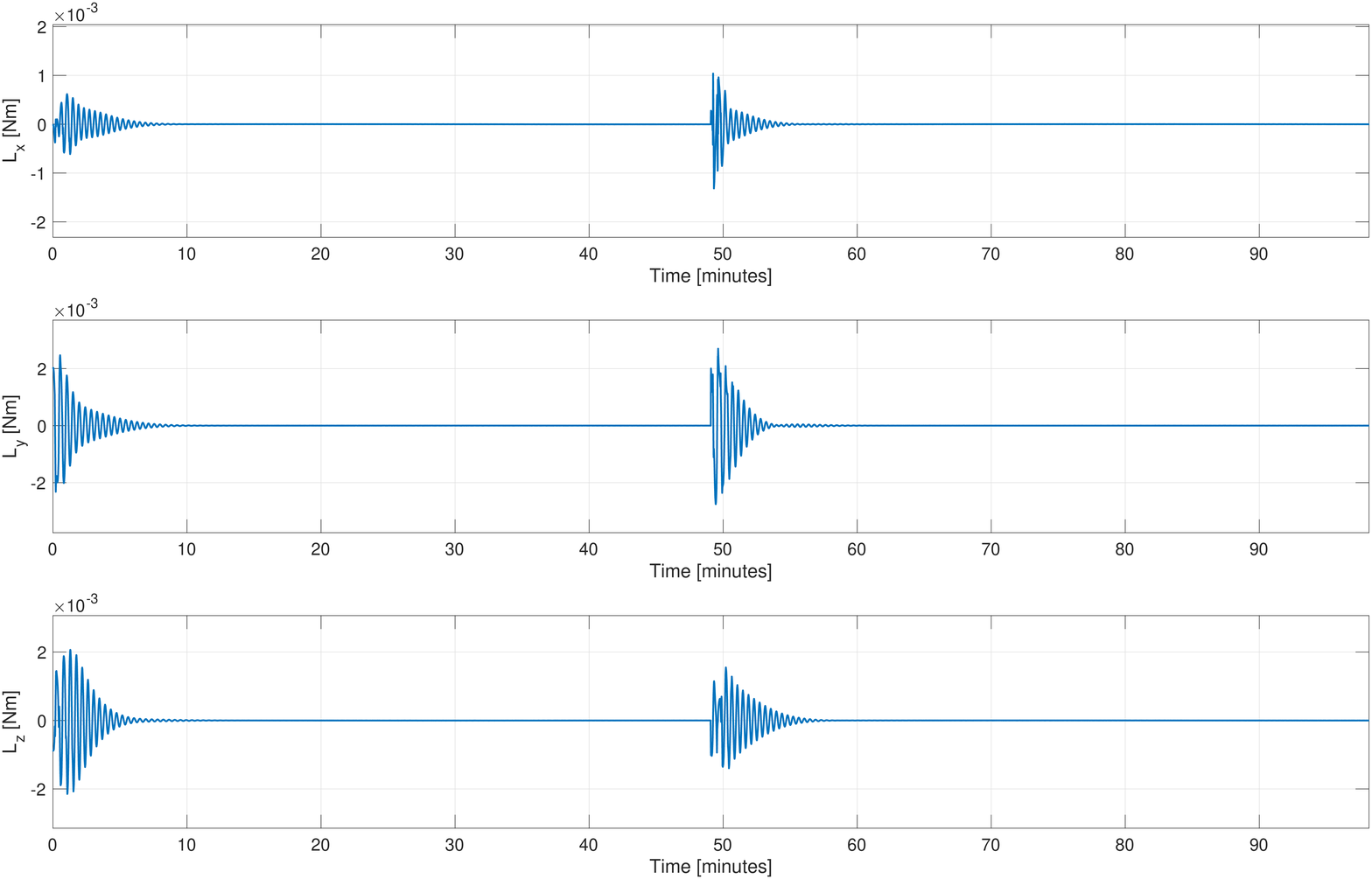}
 \caption{Slew and Tracking Simulation Results - Required Torque (Re-pointing)}
 \label{TorqueSlewRepointing}
 \end{figure}

\begin{figure}[H]
   \centering
   \includegraphics[scale=0.3]{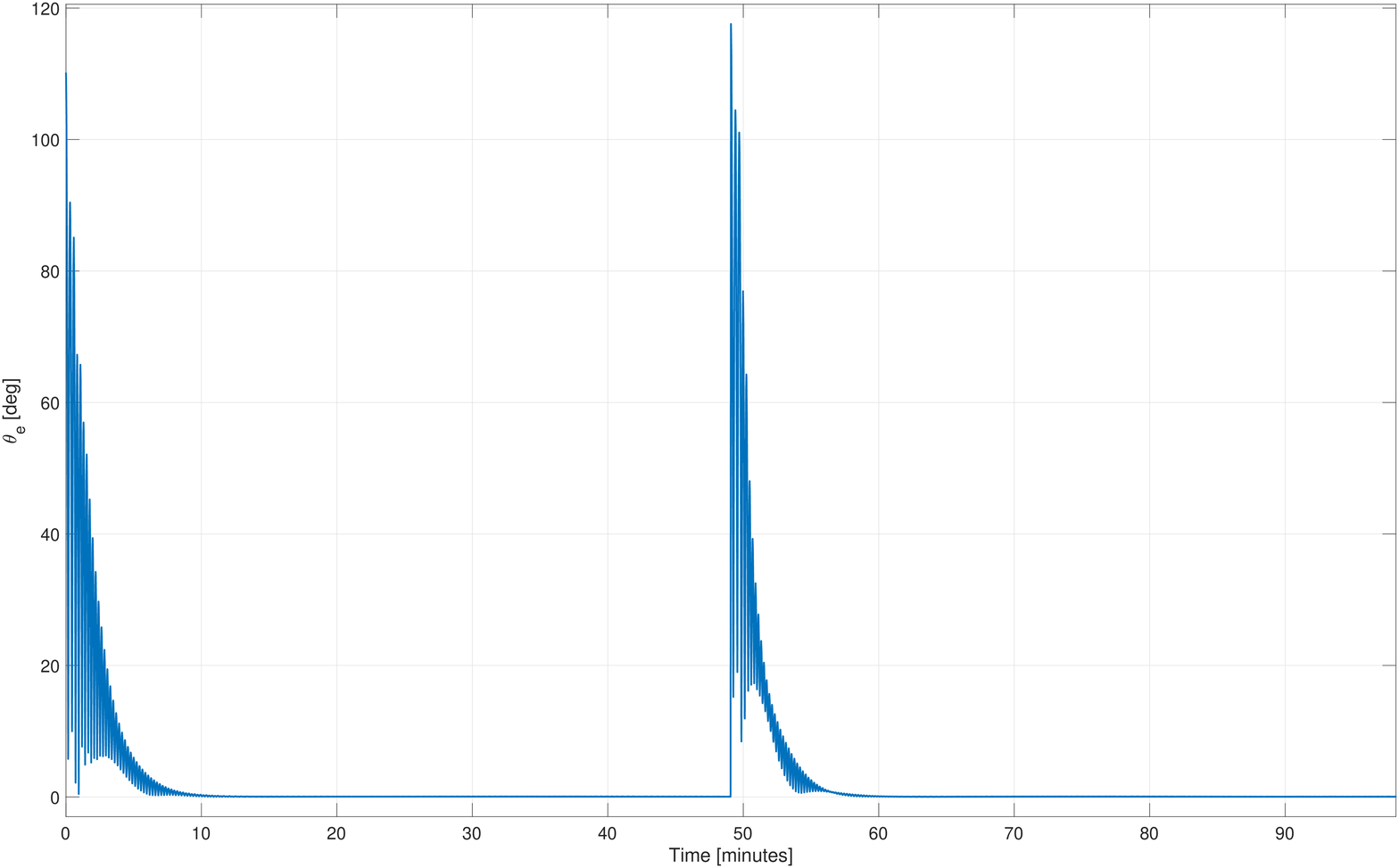}
   \caption{Slew and Tracking Simulation Results - Pointing Accuracy (Re-pointing)}
   \label{PointingAccuracyRepointing}
   \end{figure}

\begin{figure}[H]
   \centering
   \includegraphics[scale=0.3]{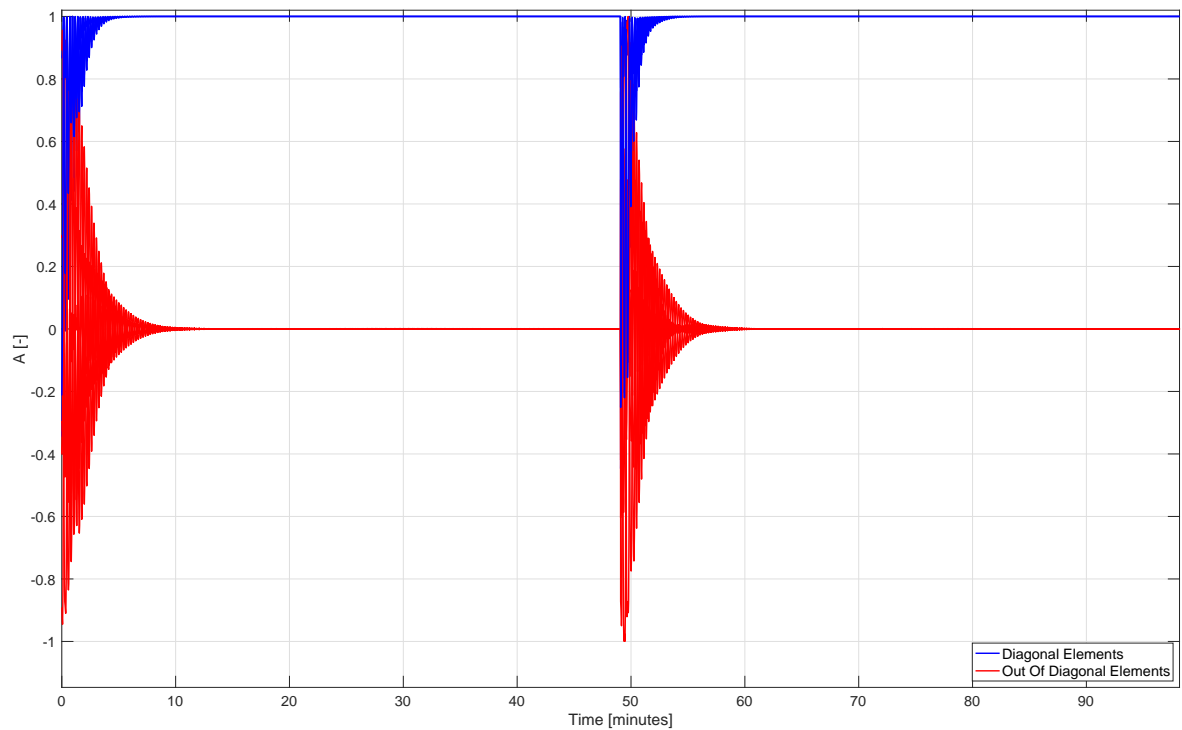}
    \caption{Slew and Tracking Simulation Results - Attitude Error (Re-pointing)}
    \label{Attitude ErrorRepointing}
   \end{figure}

 Figure~\ref{OmegaSlewRepointing} , Figure~\ref{TorqueSlewRepointing}, Figure~\ref{PointingAccuracyRepointing}  and Figure~\ref{Attitude ErrorRepointing} show the results mentioned above. 
 
 It can be appreciated how the control system is able to reorient the satellite ensuring the required pointing accuracy. Also in this case the limits imposed by the reaction wheels are respected and the attitude error converges to the identity matrix.
 
   \section{Conclusions}
Spacecraft attitude and control in short consist of a continuous  check that  a satellite points in some desired direction. But really attitude and control should be estimated simultaneously, nevertheless   they are separated to some extent due to a "separation theorem". In such a way, for the most space missions design, attitude determination dynamics can be designed separately from the control.
The case study presented here  holds almost the main theories and the state of the art, analyzing the transient performances of the control system by  solving,  by suitable numeric algorithms,  the nonlinear differential equations modeling its dynamics and stability. 
After a short description of the system features, we performed several simulations. The main concern angular velocity, dipole moments and control torques during the tumbling. Afterwards angular velocity, required torque and pointing accuracy have been monitored during slew and star tracking.
In any case we could ascertain that all the transients   behave in due way and extinguish
within the prescribed duration  according to design specifications and according to our stability analysis in the sense of Lyapunov.
The paper can be seen as  our first step on the subject: further advanced techniques of control have been taken also into account (see \cite{biggs2018attitude}, \cite{sin2021attitude}  and \cite{wang2021nonlinear}) and will be exploited in a future work.

\printbibliography
\end{document}